\documentstyle[prl,aps,epsf]{revtex}

\preprint{\bf PREPRINT}

\newcommand{\be}{\begin{equation}}
\newcommand{\ee}{\end{equation}}
\newcommand{\bea}{\begin{eqnarray}}
\newcommand{\eea}{\end{eqnarray}}
\newcommand{\eq}[1]{eq.~(\ref{#1})}

\newcommand{\Eq}[1]{Eq.~(\ref{#1})}

\begin{document}
\columnsep0.1truecm
\draft

\twocolumn[\hsize\textwidth\columnwidth\hsize\csname@twocolumnfalse%
\endcsname

\title{Statistics of Earthquakes in Simple Models of Heterogeneous Faults}
\author{Daniel S. Fisher, Karin Dahmen, Sharad Ramanathan}
\address{Lyman Laboratory of Physics,
Harvard University, Cambrigde, MA, 02138}
\author{Yehuda Ben-Zion}
\address{Department of Earth Sciences, Univ. of Southern CA,
Los Angeles, CA, 90089-0740}

\maketitle

\begin{abstract}
 
Simple models for ruptures along a heterogeneous
earthquake fault zone are studied,
focussing on the interplay between the roles of disorder and 
dynamical effects. A class of models are found to operate 
naturally at a critical point whose properties yield power law
scaling of earthquake statistics. 
Various dynamical effects can change the behavior to a distribution
of small events combined with characteristic system size events.
The studies employ various analytic methods as 
well as simulations.

\end{abstract}

]

\narrowtext

The Gutenberg-Richter law \cite{Gutenberg}
for the statistics of earthquakes --- frequency
inversely proportional to a power of the seismic moment
--- is well established
over about 10 orders of magnitude.
It is clearly a property of regional fault {\it systems}. The statistics of 
earthquakes on {\it individual} faults is much more controversial,
indeed given the degree of geometrical complexity usually observed it
is not even clear whether single faults are well defined.
Nevertheless, statistics in various narrow fault zones in which
slip is primarily along one direction --- which we will henceforth refer
to as ``faults'' --- have been studied, 
and the behavior is found to vary 
substantially.
In particular, Wesnousky\cite{Wesnousky} has found that 
faults with large total displacement which are 
relatively regular typically have a power law 
distribution only for small events --- if at all --- and events with a 
much larger characteristic size in which the whole fault slips, with 
few events in between.
In contrast, less mature faults with more
irregular geometries can have power law 
statistics over the whole range of observed magnitudes\cite{Wesnousky}.

In this paper, we will show that simple models which include fault plane 
heterogeneities can exhibit both of these types of behavior
and analyze the origin of the power law statistics and departures
from it in these systems.
In particular, we will argue that power law statistics can be 
understood quantitatively in terms of proximity to a 
specific non-equilibrium
dynamical critical point.
Like most critical points, the resulting exponents, although ``universal'',
will depend on certain properties in the system: the dimensionality,
the range of interactions, randomness, and perhaps other aspects.

Most previous work on simple models has involved variants of the 
Burridge-Knopoff (or ``sliderblock'') model in which the randomness
is generated dynamically and inertia and friction laws play an essential 
role\cite{Carlson-review}.
These systems appear to exhibit power law statistics over some range 
with a cutoff beyond some magnitude and with most of the slip occurring in 
larger, system size events.
But the understanding of the origin of the
power-law behavior is
very limited.
Our approach here will be to start with analytic understanding of a 
class of models and then add in various additional physical features
by analytic scaling arguments in the framework of the renormalization
group (RG), aided by numerical studies.

To investigate possible critical points, we first study infinite systems
driven by a constant drive force $F$. The dynamical variables
$u({\bf r},t)$ represent the
discontinuity across the fault plane in the component of the
displacement in the 
direction of slip. We consider general equations of motion of the form:
\be
\label{motion}
\eta \partial u({\bf r}, t)/ \partial t
= F + \sigma({\bf r},t) - f_R [ u({\bf r}, t),
{\bf r}, \{ u({\bf r}, t'<t) \} ]
\ee
where
\be 
\sigma({\bf r},t) = \int\limits_{-\infty}^t dt' \int d^d r'
J({\bf r}-{\bf r}',t-t')[u({\bf r}',t')
-u({\bf r},t)] 
\ee
is the stress and $f_R$ is a quenched random ``pinning'' force 
crudely representing
inhomogeneities in the friction, asperities, stepovers etc., which in 
general can depend on the local past history (e.g. as in velocity dependent
friction). The dynamics will be determined by this local history dependence,
the stress transfer function $J({\bf r},t)$, and the coefficient
$\eta$\cite{friction}.

Substantial simplifications occur if $f_R$ is history independent and
$J({\bf r},t) \geq 0 $ for all $({\bf r},t)$;
we will call these {\it monotonic} models.
Related monotonic models have been studied extensively in various other
contexts\cite{depinning,Ertas}.
Their crucial simplifying feature is that the steady state 
velocity $\overline{v} \equiv \langle \partial u/\partial t \rangle $
is a history independent function of $F$\cite{Middleton-nopassing}.
For $F$ less than a critical force $F_c$, $\overline{v} = 0$,
while  just above $F_c$, $\overline{v} \sim (F-F_c)^\beta$.
Universal scaling behavior exists on large length scales near $F_c$.
Quasi-static properties such as exponents and scaling functions
depend on only a few quantities: the spatial dimension 
$d$; the range
of the interactions if they are long-range, {\it i.e.} with the 
static stress transfer $J_s(r) \equiv \int dt J(r, t) \sim 1/r^{d+\Gamma}$,
with $\Gamma<2$; and the range of correlations in $f_R$, which we will 
generally assume are short-range in $u$ and ${\bf r}$.
Long time dynamic properties such as $\beta$ depend in addition 
on the small $\omega$ dependence of $J({\bf q}, \omega) $.
If $F$ is adiabatically increased towards $F_c$, the system 
moves from one metastable configuration to another by a sequence of 
``quakes'' of various sizes. 
The ``quakes'' can be characterized 
by their radius $R$, the $d$-dimensional area $A$ which slips (by more 
than some small cutoff), their moment $M \equiv \int d^d {\bf r} 
\Delta u ({\bf r})$, a typical displacement $\Delta u \sim M/A $,
and a duration $\tau$.

From RG expansions\cite{depinning}
around a dynamic mean field 
version of \Eq{motion} and scaling arguments it is found that 
for large quakes $\Delta u \sim R^\zeta $, $A \sim R^{d_f}$ with $d_f \leq d$ 
a fractal dimension, $M \sim R^{ {d_f}+ \zeta }$ and $\tau \sim R^z$.
The distribution of moments is 
\be
\label{scaling}
P(M) dM \sim dM/M^{1+B} \rho_\infty (M/\hat{M}) 
\ee
with $\rho_\infty$ a universal scaling function which decays exponentially
for large argument.
The cutoff $\hat M$ for large moments is characterized by a 
correlation length ---
the largest likely radius --- $ \xi \sim 1/(F_c-F)^\nu $
with $\hat{M} \sim \xi^{d_f+\zeta}$.
In mean-field theory, $B=1/2$, the quakes are fractal and 
displacements are of order the range of correlations in $f_R(u)$, 
{\it i.e.} $\zeta=0$.
The mean-field exponents are valid for $d> d_c(\tilde\Gamma)=2 \tilde\Gamma$
where $\tilde\Gamma \equiv {\it min}(\Gamma,2)$\cite{Ertas}.
For a planar fault in an elastic half space,
$d=2$ and $\Gamma=1$\cite{Yehuda}; 
the {\em physical} system is thus {\it at the upper 
critical dimension} $d=d_c$\cite{one-dim}.

As usual, at the upper critical dimension, there are logarithmic 
corrections to mean-field results. 
We find barely fractal quakes with $A\sim R^2/ \ln R $
so that the fraction of the area slipped decreases only as $1/\ln r$
away from the ``hypocenter''. The typical slip is 
$\Delta u \sim ( \ln R )^{1/3}$ so that 
$M \sim R^2/(\ln R)^{2/3} $.
The scaling form of $P(M)$ is the same as \Eq{scaling}
with the mean-field $\rho_\infty$, although for $M << \hat{M}$,
$ P(M) \sim (\ln M )^{1/3}/ M^{3/2}$
so that $B$ will be virtually indistinguishable from $1/2$.

We now consider more realistic drive and finite-fault-size
effects. Driving the fault by very slow motion 
far away from the fault is 
roughly equivalent to driving it with a weak spring,
{\it i.e.} replacing $F$ in \Eq{motion} by
$F({\bf r}, t)= K [ vt-u({\bf r}, t)] $. With 
$v \rightarrow 0$ 
the system must then operate with the spring stretched to make 
$F({\bf r}, t) \lesssim F_c $ at least on average;
it will actually operate just below $F_c$.
Under a small increase, $\Delta F$, with constant force drive,
$\langle \Delta u \rangle \approx n \Delta F \int M P(M) dM $
with $n$ the number of quakes per unit area per increase in $\Delta F$;
$n(F)$ is non-singular at $F_c$\cite{depinning}.
The known scaling laws yield 
$\langle \Delta u \rangle \sim \Delta F \xi^{(2 \tilde\Gamma + \zeta)(1-B)}
\sim \Delta F \xi$ for our case.
For consistency, we must have in steady state with the spring drive,
$ K v \Delta t = \Delta F = K \Delta u $
so that the system will operate with a correlation length 
$\xi \sim 1/K^{1/\tilde \Gamma} $, {\it i.e.} $1/K$ for our case.
For a fault section with linear dimensions of order $L$, drive either 
from uniformly moving fault 
boundaries or from a distance $\sim L$ perpendicularly away from 
the fault plane will be like $K\sim 1/L$ so that
the power-law quake distribution will  
extend out to roughly the
system size $\xi \sim L$.
For smaller quakes, {\it i.e.} $R << L$, the behavior will be the 
same as in the infinite system with constant $F$ drive, but the cutoff of 
the distribution of moments will be like \Eq{scaling} 
with a different cutoff function $\rho$ that depends on the shape 
of the fault, how it is driven, and the boundary conditions.

\begin{figure}
\narrowtext \epsfxsize=2.9truein \vbox{\hskip 0.15truein
\epsffile{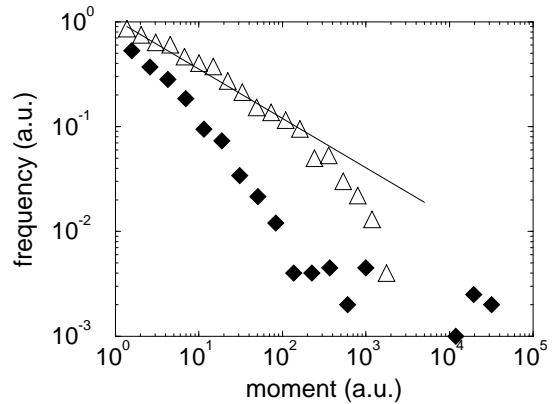}}
\medskip
\caption{Histograms of moments for a simulation of a 
rectangular fault with 32x128 cells for the discrete monotonic 
quasistatic model. Triangles: without dynamical weakening ($\epsilon=0$).
Diamonds: with dynamic weakening with $\epsilon=0.95$.
($\epsilon$ is defined in 
\protect{\eq{weakening}}.)
The straight line indicates the 
predicted slope $B=1/2$. 
\label{moments}}
\end{figure}

\begin{figure}
\narrowtext \epsfxsize=2.9truein \vbox{\hskip 0.15truein
\epsffile{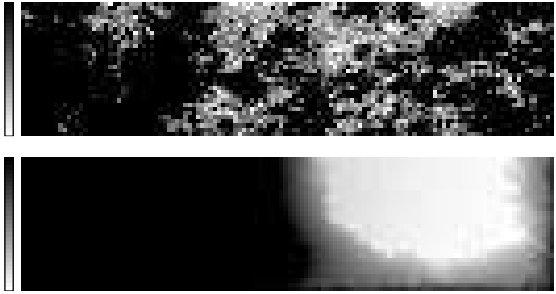}}
\medskip
\caption{Distribution of horizontal slip, $u$, along a fault 
with 32x128 cells for a {\it single} large quake event.
The lighter a cell the bigger its slip during the quake.
Top (a) almost fractal quake with a total moment of 1750 (and 1691 
cells 
failing) for the monotonic model
without any dynamical effects ($\epsilon=0$).
Bottom (b) ``cracklike'' quake with a total moment 
of 16922 (and 2095 cells
failing) for the model
with dynamic weakening ($\epsilon=0.95$).
In both cases the system is driven by horizontally 
creeping fault boundaries (sides and bottom) while 
the top boundary is free.
\label{faults}}
\end{figure}

We have tested these conclusions numerically by studying a discrete space, 
time, and displacement version of a monotonic \Eq{motion} with 
quasistatic stress
transfer appropriate to an elastic half space\cite{Yehuda}. 
The slip, $u$,
is purely in the horizontal direction along the fault and
$f_R[u({\bf r})]$ is a series of equal height spikes with 
spacings  
which are a random function of ${\bf r}$.
When $\sigma({\bf r},t) > f_R[u({\bf r},t]$, $u({\bf r})$ jumps to the 
next spike. 
The boundary conditions on the bottom and sides are uniform 
slip ---
($u=vt$) with infinitesimal $v$ --- and stress free on the top.
The statistics of the moments of the quakes are shown by the triangles
in Fig.~\ref{moments}.
Although the uncertainties are appreciable, relatively good agreement 
is found with the prediction $B=1/2$.
A typical large quake 
is illustrated in Fig.~\ref{faults}~(a);
it appears almost fractal as predicted and will tend to stay away
from the bottom and sides.
The ratios of the moments of quakes 
to their areas have been studied and found 
to grow only very slowly with the area, as predicted.
This is in striking contrast to earthquakes in conventional crack models
which are compact and have $\Delta u \sim R $ ({\it i.e.} $\zeta=1$), 
so that $M/A \sim \sqrt{A}$.

Because the system is at its critical dimension, the cutoff function $\rho$
of the moment distribution appropriate to the boundary conditions,
as well as various aspects of the shapes and dynamics of quakes 
can be computed.
For quasistatic stress transfer, $J({\bf r},t) \sim \delta(t)/r^3 $,
in the infinite system, the quake durations are found to 
be $\tau \sim R^z$ with $z< 1$ for $d<d_c$,
corresponding to an unphysical supersonic propagation of 
disturbances\cite{Ertas}.
In the marginal dimension $d=d_c$, $z=1$ with logarithmic corrections.
A more physical dynamics with sound-travel-time delay has  
slower growth of the quakes with $z=1$ in all dimensions.
In either case the growth will be very irregular --- including regions
starting and stopping --- in contrast to crack models and what is often 
assumed in seismological analyses of earthquakes.
A crucial feature of monotonic models is that the slip profile
$\Delta u({\bf r})$ of a quake is {\it independent of
the dynamics}\cite{Middleton-nopassing}. 
But the most interesting dynamical issues concern the
effects left out of the monotonic models that can make this feature 
breakdown. We first consider including some weakening effects
of sections which have already slipped in a given quake. 
This is best studied in the discrete model.
To crudely model a difference between static and dynamic friction 
we choose,
\be
\label{weakening}
f_R = \tilde f_R [ u({\bf r}),{\bf r}] 
\{1 - \epsilon 
\Theta[u({\bf r}, t)-u({\bf r}, t-T)] \}
\ee
with $T$ a cutoff time much longer than the duration of the largest
quakes, but much smaller than the interval between the quakes.
This can crudely represent a difference between static and dynamic
friction.
For $ \epsilon =0 $ and non-negative $J$ the model is monotonic, while 
for $\epsilon > 0$ it is non-monotonic.
The effects of small weakening can be analyzed perturbatively.

With $\epsilon=0$, consider a quake of diameter $R_1$ ($ << L$ {\it or}
$\xi $),
with moment $M_1$ and area $A_1$: {\it i.e.} $A_1$ sites have slipped. 
If a small $\epsilon$ is turned on at the end of the quake, all slipped
sites that are within $\epsilon$ of slipping will now slip again ---
this will be $N^{ex}_2 \sim \epsilon A_1 $ sites. The simplest 
justifiable guess is that each of these will cause an approximately 
independent secondary quake.
The total moment of these secondary quakes will be dominated by the 
largest one, so the extra moment 
will be $M^{ex}_2 \sim (\epsilon A_1)^{1/B} $.
If $ M^{ex}_2 << M_1 $ this process can continue but will not increase
the total moment substantially. 
But if $M^{ex}_2 \sim M_1 $, the process can continue with a larger area 
$A_2$ and hence a larger $M^{ex}$, leading 
to runaway.
The scaling laws yield $B= ( d_f-\tilde\Gamma + \zeta)/(d_f + \zeta) $
and $ A \sim M^{d_f/(d_f+\zeta)} $ with $\zeta < \tilde \Gamma $,
so that for any $\epsilon$, for large enough $M_1$, $M_1 \gtrsim M_D 
\sim \epsilon^{-(d_f+\zeta)/(\tilde\Gamma-\zeta)} $,
$M^{ex}_2$ will be comparable to $M_1$ and the
quake will become much larger.
In the case of interest $M_D \sim \epsilon^{-2}$.
In the force driven infinite system for $F \lesssim F_c $,
quakes of size $\xi $ will runaway and become infinite if 
$\xi > \epsilon^{-1/(\tilde \Gamma - \zeta )} $.
Since $\xi \sim (F-F_c)^{-\nu}$ and $1/\nu = \tilde\Gamma - \zeta$,
this will occur for $F_c - F < C_w \epsilon $ with some constant $C_w$.
This result is very intuitive and justifies a posteriori the 
assumptions leading to it:
Since on slipping, the random pinning forces, $f_R$ in a region 
are reduced by order $\epsilon$,
the effective critical force $F_c$ for continuous slip will have been 
reduced by order $\epsilon$; thus if $F > F_c(\epsilon)=F_c -C_w \epsilon$, 
the mean velocity $\overline v$ will be nonzero.

\begin{figure}
\narrowtext \epsfxsize=2.9truein \vbox{\hskip 0.15truein
\epsffile{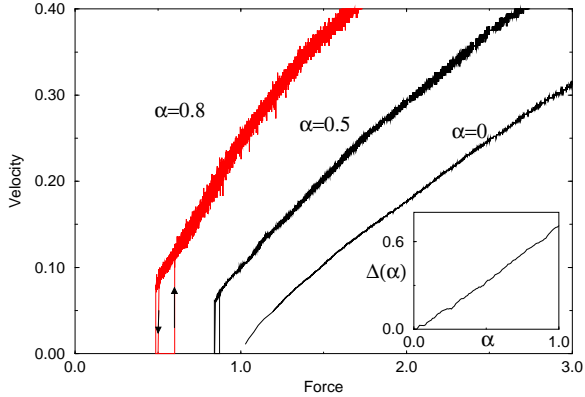}}
\medskip
\caption{Mean velocity versus force for one dimensional system with a 
non-monotonic kernel $J(x,t)=\delta(t-x)/x^2 +\alpha\delta{'}(t-x)/x$ for
$\alpha=0.8,0.5,0$.  A spring or boundary loaded system 
will traverse
the hysteresis loops in the direction indicated. Inset: the threshold
force, $F_c^{\uparrow}(\alpha)$, on increasing the load;
$\Delta_{\alpha}=[1-F_{c}^{\uparrow}(\alpha)/
F_{c}^{\uparrow}(\alpha=0)]^{1/2}$
is plotted vs. $\alpha$.
\label{force-driven}}
\end{figure}

A similar but more subtle effect can be caused by stress pulses
that result from non-positive $J({\bf r},t)$; these arise naturally when 
one includes elastodynamic effects.
We consider
\be
\label{coupling}
J({\bf r}, t) \sim 
 \delta(t-r/c)/r^{d+\Gamma} + \alpha \delta'(t-r/c)/c r^{d+\gamma} 
\ee
with $c$ the sound speed.
The scalar approximation to elasticity in a half space 
corresponds to $d=2$, $\Gamma=1$, $\gamma=0$, and $\alpha=1$.
If a region slips forward, the stress at another point first has a 
short pulse at the sound arrival time from the second term 
in \Eq{coupling}, then settles down to its smaller static value,
{\it i.e.} it is non-monotonic.
The magnitude of these stress pulses and their duration is set by 
various aspects of the models, for example 
larger $\eta$ in \Eq{motion} implies weaker stress pulses
as the local motion will be slower.
By considering which of the sites in a long quake with 
$\alpha=0$
can be caused to slip further by such stress pulses 
--- here the dynamics matters --- we find that
runaway will occur for $M \geq M_D \sim \alpha^{-4} $
for the physical case. 
We have checked this in $d=1$ with $\Gamma =1$ and $\gamma=0$
finding the predicted reduced critical force 
$F_c(\alpha) \sim  F_c - C_p \alpha^2 $
as shown in Fig.~\ref{force-driven}.
These 1-d simulations also reveal a hysteretic
$\overline{v}(F)$ curve in finite systems.
This should also occur with the velocity weakening model.

We can now understand what should happen with either weakening or stress
pulses in finite systems driven with a weak spring
or with slowly moving boundaries. 
As the system is loaded, quakes of increasing
size can occur. If the system is small enough that it cannot sustain quakes
with $M>M_D(\epsilon,\alpha)$ then the behavior will not be much different 
from the monotonic case with $\epsilon = \alpha = 0 $.
This will occur if $ L < R_D(\epsilon, \alpha) \sim M_D^{1/2} \sim 
{\it max}(C_\alpha/\alpha^2, C_\epsilon/\epsilon) $
with appropriate coefficients $C_\alpha$, $C_\epsilon$,
which will depend on the amount of randomness in the fault. 
But if $L>R_D$, quakes of size of order $R_D$ will runaway and most of the
system will slip, stopping only when the load has decreased enough to 
make the loading forces less than the {\it lower} end of the 
hysteresis loop in ${\overline v}(F)$ (as in Fig.~\ref{force-driven}).
Because of the tendency of regions that have already slipped
to slip further, and the consequent buildup of larger stresses near
the boundaries of the slipped regions, large events in systems with 
dynamic weakening will be much more cracklike than in monotonic models,
probably with $\Delta u \sim L $.
Statistics of quakes with weakening, $\epsilon$, reasonably large,
but no stress pulses ($\alpha = 0$) are shown in Fig.~\ref{moments}
and in \cite{Yehuda};
note the absence of 
quakes with intermediate moments.  A typical large event in this case is 
shown in Fig.~\ref{faults}~(b); it appears to be crack-like.

In this paper we have shown that simple models of heterogeneous 
faults --- with the dimensionality and long-range elastic interactions
properly included --- can give rise to either power-law statistics
of earthquake moments or a distribution of small events combined with 
characteristic system size events. Which behavior --- or intermediate
behavior --- obtains is found to depend on a number of physical 
properties such as frictional weakening and dynamic stress transfer,
analogs of which should definitely exist in real systems.
In the power-law-regime the conventionally defined Gutenberg-Richter
exponent $b \equiv 3 B/2 $ is is found to be $b=3/4$.
This is intriguingly close to values observed by Wesnousky\cite{Wesnousky},
but it 
is not clear if any significance should be attached to this.

More significant is the framework that we have built,
which enables certain results (and many more not presented here)
to be obtained analytically and others to be understood by scaling 
arguments. It is hoped that other physical phenomena such as
geometrical disorder, side branching, and 
multiple cracks might start to be addressed in this framework.
We note one extra effect which can be readily 
analyzed: long range correlations in the randomness (perhaps caused
by prior history of the fault). Varying the power-law of the decay 
of 
correlations of $f_R$ increases $B$ continually from $1/2$ to 
$2/3$ and $\zeta$ concomitantly from 0 to 1 with the quakes becoming more 
compact and crack-like as the randomness correlations become longer
range.

We would like to thank Jim Rice, Deniz Ertas, Chris Myers, and Jim Sethna
for many useful conversations. 
This work is supported in part by the Harvard Society
of Fellows and by NSF via DMR 9106237, 9630064 and Harvard's MRSEC.


\begin{references}
\bibitem{friction} Linear velocity-strengthening friction can be 
absorbed into $\eta$.
\bibitem{one-dim}
In contrast, for very large quakes that 
rupture through the whole depth of the crust, the fault becomes 
essentially one dimensional with $\Gamma=2$, although the different 
boundary conditions etc. will make our analysis not applicable.
\end{references}
\end{document}